# Memetic Elitist Pareto Evolutionary Algorithm for Virtual Network Embedding

Ashraf A. Shahin[1,2]

[1]College of Computer and Information Sciences, Al Imam Mohammad Ibn Saud Islamic University (IMSIU) Riyadh, Kingdom of Saudi Arabia

[2]Department of Computer and Information Sciences, Institute of Statistical Studies & Research, Cairo University, Egypt

Correspondence: Ashraf A. Shahin, College of Computer and Information Sciences, Al Imam Mohammad Ibn Saud Islamic University (IMSIU) Riyadh, Kingdom of Saudi Arabia. E-mail: ashraf_shahen@ccis.imamu.edu.sa

**Abstract**

Assigning virtual network resources to physical network components, called Virtual Network Embedding, is a major challenge in cloud computing platforms. In this paper, we propose a memetic elitist pareto evolutionary algorithm for virtual network embedding problem, which is called MEPE-VNE. MEPE-VNE applies a non-dominated sorting-based multi-objective evolutionary algorithm, called NSGA-II, to reduce computational complexity of constructing a hierarchy of non-dominated Pareto fronts and assign a rank value to each virtual network embedding solution based on its dominance level and crowding distance value. Local search is applied to enhance virtual network embedding solutions and speed up convergence of the proposed algorithm. To reduce loss of good solutions, MEPE-VNE ensures elitism by passing virtual network embedding solutions with best fitness values to next generation. Performance of the proposed algorithm is evaluated and compared with existing algorithms using extensive simulations, which show that the proposed algorithm improves virtual network embedding by increasing acceptance ratio and revenue while decreasing the cost incurred by substrate network.

## 1. Introduction

Network virtualization plays an important role in cloud data centers by allowing multiple simultaneous virtual networks (VNs) to share single substrate network (SN). Network virtualization increases revenue of cloud data centers and improves utilization of physical resources by increasing number of accommodated virtual networks.

However, mapping virtual networks' resources to physical network's resources is known to be nondeterministic polynomial-time hard (NP-hard) even if all incoming requests are previously known (Cheng et al., 2012). This problem is usually referred to as the *Virtual Network Embedding* (VNE) problem (Zhang et al., 2013). Figure 1 shows an example of VNE. VNE solution maps each virtual node to a substrate node and each virtual link to a loop free substrate path, which consists of substrate links. For convenience, in the rest of this paper, we do not differentiate between virtual network embedding and virtual network mapping.

In the last few years, many studies have proposed several optimal techniques to solve small instances of the problem in reasonable time (Lischka & Karl, 2009; Chowdhury et al., 2009; Chowdhury et al., 2012; Cheng et al., 2011). However, current cloud data centers contain thousands of servers and using traditional optimal techniques to find the optimal solution in cloud data centers becomes unaffordable.

Therefore, metaheuristic-based techniques have received more attention to be used for finding near optimal solution in short execution time (Zhang et al., 2013; Cheng et al., 2012; Yuan et al., 2013; Wang et al., 2013). Genetic algorithm (GA) is one of the widely used metaheuristic-based techniques. Genetic algorithms have been successfully applied to many optimization problems in different areas. However, genetic algorithms have some drawbacks like the possible premature convergence. Due to the loss of diversity within the population, GAs can converge even though the near optimal solution is not yet found.

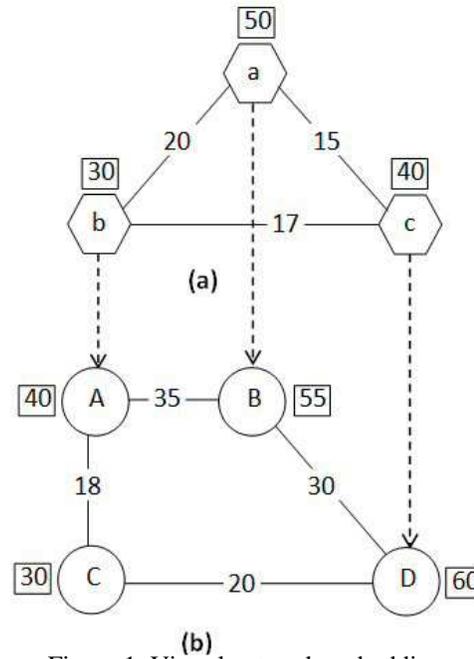

Figure 1. Virtual network embedding example

A memetic algorithm is an extension of the traditional genetic algorithm to reduce the problem of premature convergence. Memetic algorithms are a combination of population-based global search and heuristic-based local search. Global search finds the near global optimal and local search finds the best solution. Another important factor that affects the acceptance ratio of virtual network requests is substrate resources fragmentation. Substrate resources become fragmented due to the dynamic arrival and departure of virtual network requests over time.

In this paper, we propose a memetic elitist pareto evolutionary algorithm for embedding virtual networks (*MEPE-VNE*). *MEPE-VNE* reduces the loss of good solutions by employing elitist selection. Elitist selection passes a limited number of individuals with the best fitness values to the next generation without applying crossover or mutation operations on them. *MEPE-VNE* combines Pareto-based multi-objective algorithms with local search to optimize virtual network embedding. The set of Pareto optimal front solutions is gained using non-dominated sorting genetic algorithm-II (NSGA-II), which is one of the most famous Pareto optimal solution algorithms (Deb et al., 2002). The performance of the proposed algorithm is evaluated and compared with some of existing algorithms using extensive simulations, which show that the proposed algorithm outperform existing algorithms by improving virtual networks' acceptance ratio and revenue.

The rest of this paper is organized as follows. Section 2 gives a brief explanation of the key basic concepts of multi-objective optimization. Section 3 gives a short overview of related work. Section 4 presents virtual network embedding model and problem formulation. Section 5 describes the proposed algorithm. Section 6 evaluates the proposed virtual network-embedding algorithm using extensive simulations. Finally, in Section 7 we conclude this paper.

## 2. Background

*Multi-objective optimization*: multi-objective optimization problem can be depicted as following (Reyes-sierra and Carlos, 2006):

$$\text{Minimize } \vec{f}(\vec{x}) = \big(f_1(\vec{x}), f_2(\vec{x}), \ldots, f_k(\vec{x})\big)$$

subject to:

$$g_i(\vec{x}) \leq 0 \quad i = 1, 2, \ldots, m$$
$$h_i(\vec{x}) = 0 \quad i = 1, 2, \ldots, p$$

Where $\vec{x} = (x_1, x_2, \ldots, x_n) \epsilon \vec{X} \subset \mathbb{R}^n$ is a decision vector consists of $n$ decision variables, $\vec{X}$ is a decision space, $\vec{f}(\vec{x}) = (f_1(\vec{x}), f_2(\vec{x}), \ldots, f_k(\vec{x})) \epsilon \vec{Y} \subset \mathbb{R}^k$, is an objective vector consists of $k$ objective functions, $\vec{Y}$ is an objective space, $f_i(\vec{x}): \mathbb{R}^n \to \mathbb{R}, i = 1,2, \ldots, k$ are objective functions, $g_i(\vec{x}): \mathbb{R}^n \to \mathbb{R}, i = 1,2, \ldots, m$, $h_i(\vec{x}): \mathbb{R}^n \to \mathbb{R}, i = 1,2, \ldots, p$ are inequality and equality constraints functions of the problem. Multi-objective optimization problem tries to find decision vector $\vec{x}$ in decision space $\vec{X}$ that will optimize objective vector $\vec{f}(\vec{x})$.

*Definition 1 (Feasible Solution Set).* Feasible solution set is the set of all decisions from a decision space $\vec{X}$ that satisfy all inequality and equality constraints. Feasible solution set is denoted by $\vec{X_f}$, where $\vec{X_f} \subset \vec{X}$.

*Definition 2 (Pareto dominance).* Let $\vec{x^1}, \vec{x^2} \epsilon \vec{X_f}$, we say that $\vec{x^1}$ dominates $\vec{x^2}$ (denoted by $\vec{x^1} \prec \vec{x^2}$) or $\vec{x^2}$ is dominated by $\vec{x^1}$ iff $f_i(\vec{x^1}) \leq f_i(\vec{x^2}) \ \forall \ i \in \{1,2,\ldots,k\} \land \exists \ i \in \{1,2,\ldots,k\} : f_i(\vec{x^1}) < f_i(\vec{x^2})$

*Definition 3 (Non-domination).* We say that a decision vector $\vec{x} \epsilon \vec{X}$ is non-dominated with respect to $\vec{X}$, if $\nexists \ \vec{x'} \epsilon \vec{X}$ such that $\vec{x'} \prec \vec{x}$.

*Definition 4 (Pareto optimality).* A decision vector $\vec{x^*} \epsilon \vec{X_f}$ is Pareto optimal if $\vec{x^*}$ is non-dominated with respect to $\vec{X_f}$.

*Definition 5 (Pareto optimal set).* The set of all Pareto optimal decision vectors is called Pareto optimal set and is denoted by $\mathcal{P}^*$.

*Definition 6 (Pareto front).* The Pareto front $\mathcal{PF}^*$ is defined by: $\mathcal{PF}^* = \{ \vec{f}(\vec{x}) \mid \vec{x} \in \mathcal{P}^*\}$

**3. Related work**

Solving VNE problem is NP-hard even if all incoming requests are previously known. Even if all virtual nodes are mapped, finding the optimal virtual links mapping is still NP-hard. Therefore, optimal solution can only be found for small problem instances (Fischer et al., 2013b). Consequently, several heuristic algorithms have been proposed to find a near optimal solution (Zhang et al., 2012; Sun et al., 2013; Ghribi et al., 2013; Beloglazov & Buyya, 2012; Nogueira et al., 2011). Some algorithms have been proposed to find optimal VNE solutions to be used as optimal bound for the heuristic based VNE solutions (Cheng et al., 2011; Botero et al., 2012). Shahin (2015b) proposed two VNE algorithms to find optimal VNE solutions. The proposed algorithms increase the acceptability of fragmented substrate networks by embedding virtual networks on best-fit sub-substrate networks. Substrate networks are fragmented due to dynamic arrival and departure of virtual network requests over time. Fragmentation of substrate networks reduces the possibility of accepting more virtual network requests in the future. The proposed algorithms are one stage, online, and backtracking algorithms. Both algorithms exploit the virtualization technology to consolidated more than one virtual node from the same virtual network to one substrate node. The second algorithm coarsens VNs using *Heavy Edge Matching* (HEM) technique to minimize the total required bandwidth. In (Shahin, 2015a), Shahin proposed another VNE technique, which coarsens virtual network using *Heavy Clique* matching technique and optimizes the coarsened virtual networks using a refined *Kernighan-Lin* algorithm.

Zhu and Ammar (2006) proposed two VNE algorithms to find exact VNE solutions. In the first algorithm, allocated substrate resources are fixed throughout the VN lifetime. To reduce the required computational to embed VNs, Zhu and Ammar divided each VN into a number of connected sub-VNs. In the second algorithm, allocated substrate resources are reallocated to increase the utilization of the underlying substrate resources. Substrate resources reallocations are prioritized for the most important VNs. Selective scheme is proposed to select the required substrate resources reallocations after prioritizing the required reallocations for the most important VNs. However, the proposed algorithms deal only with offline version of VNE problem, where all VNRs are previously known and do not deal with VNRs that dynamically arrive over time.

Lischka and Karl (2009) proposed online one stage VNE algorithm, which deals with dynamic arrival of VNRs and maps virtual nodes and virtual links during the same stage. The proposed algorithm maps VN to substrate network by finding an isomorphic subgraph inside the substrate network that achieves the CPU and connectivity requirements for the VN. During nodes mapping process, virtual nodes are sorted in descending order based on its required CPU and mapped sequentially to substrate nodes. Although, allowing multiple virtual nodes from the same

VN to be coexisted on the same substrate node leverages the advantage of ram data switch between virtual nodes coexisted on same substrate node instead of using substrate link bandwidth, the proposed algorithm prevents virtual nodes from the same VN to be co-resident on the same substrate node. Lischka and Karl incrementally increase the maximum hop limit to find the shortest substrate path that is required to embed each virtual link. However, the computational complexity of the proposed algorithm is high due to multiple operations.

To reduce complexity of the proposed algorithm in (Lischka & Karl, 2009), Di et al. (2010) sorted virtual nodes and specified the maximum hop limit based on virtual links mapping cost. Fischer et al. (2013a) proposed another improvement for the proposed algorithm in (Lischka & Karl, 2009). Fischer et al. considered energy efficiency during nodes and links mapping and allowed several virtual nodes of the same virtual network to be coexisted on the same substrate node. However, they did not take into account the mapping cost.

Yuan et al. (2013) modeled the VNE problem as an integer linear programming and proposed VNE algorithm based on discrete particle swarm optimization to solve the proposed model. The proposed algorithm allows more the one virtual node to be coexist in the same substrate node to leverage the advantages of inter ram switch techniques. Wang et al. (2013) proposed VNE algorithm similar to the proposed algorithm in (Yuan et al., 2013). Furthermore, Wang et al. presented an enhanced position assign method to make the solution space more narrow and increase the number of consolidated virtual nodes.

Cheng et al. (2011) proposed topology-aware node ranking technique, called *NodeRank*, to reflect resources and quality of connections of a node. *NodeRank* technique is inspired by *PageRank* technique used by Google's search engine, which ranks web pages to measure its popularity by applying the *Markov Random Walk (RW)* model. Using the proposed *NodeRank* technique, Cheng et al. proposed two VNE algorithms: *RW-MaxMatch*, and *RW-BFS*. *RW-MaxMatch* is two-stage VN embedding algorithm, which maps virtual nodes according to their ranks and maps virtual links between the mapped nodes to shortest substrate paths. However, *RW-MaxMatch* algorithm maps nodes without taking into consideration its relation to the link mapping, which leads to high SN's resources consumption rate. This is due to mapping neighboring virtual nodes widely separated in the SN. *RW-BFS* algorithm is a backtracking one-stage VN embedding algorithm based on breadth-first Search. *RW-BFS* improves the coordination between nodes and links mapping by embedding nodes and links at the same stage.

Zhang et al. (2013) proposed two VN embedding models: an integer linear programming model and a mixed integer-programming model. To solve these models, Zhang et al. proposed a discrete particle swarm optimization based VNE algorithm, called *RW–PSO*, to find a near optimal VNE solutions. Zhang et al. used *RW–PSO* algorithm only to propose a possible virtual node mapping solutions and mapped virtual links using shortest path algorithm and greedy k-shortest paths algorithm. Cheng et al. (2012) proposed an enhanced version of the *RW–PSO* algorithm by combining the topology-aware node ranking measure proposed in (Cheng et al., 2011) with discrete Particle Swarm Optimization VNE algorithm.

Chowdhury et al., (2009) (2012) formulated the VNE problem as a mixed integer-programming model and relaxed the proposed model to linear-programming model. To solve the proposed models with better coordination between nodes mapping stage and links mapping stage, Chowdhury et al. proposed two VNE algorithms: *D-ViNE* (deterministic VNE algorithm) and *R-ViNE* (randomized VNE algorithm).

## 4. Virtual network embedding model and problem formulation

*Substrate network (SN)*: as in our previous work (Shahin, 2015a; Shahin, 2015b), SN is modeled as a graph $G_s = (N_s, L_s)$, which is a weighted undirected graph. $N_s$ is the set of all substrate nodes and $L_s$ is the set of all substrate links. Substrate nodes are weighted by available CPU, and substrate links are weighted by available bandwidth. Figure 1(b) shows substrate network example. The numbers in rectangles represent CPU capacities and the numbers over links represent bandwidth capacities.

*Virtual network (VN):* virtual network $VN_i$ is represented as a weighted undirected graph $G_{v_i} = (N_{v_i}, L_{v_i})$. The set $N_{v_i}$ contains all virtual nodes and each virtual node is weighted by the required CPU. The set $L_{v_i}$ contains all virtual links and each virtual link is weighted by required bandwidth. Figure 1(a) shows an example of virtual network, where numbers in rectangles represent required CPU and the numbers over links represent required bandwidth.

*Virtual network requests (VNR):* virtual network request $vnr \in VNR$ is modeled as $(G_v, t_a, t_l)$, where $G_v$ is the graph of the required virtual network, $t_a$ is the arrival time, and $t_l$ is the lifetime. Substrate CPU and BW are allocated to accommodate virtual network requests. If substrate network does not have enough substrate resources to achieve virtual network request, virtual network request is rejected. Allocated substrate resources are released at the end of the lifetime.

*Virtual Network Embedding (VNE)*: virtual network embedding maps each virtual node to a substrate node and each virtual link to a loop-free substrate path. Embedding $VN_i$ on SN can be defined as $M: G_{v_i} \rightarrow (N_s', P_s')$, where $N_s' \subseteq N_s$, $P_s' \subseteq P_s$, and $P_s$ represents the set of loop free substrate paths in $G_s$.

*Virtual Network Embedding Revenue:* as in (Shahin, 2015a; Shahin, 2015b; Cheng et al., 2011; Zhang et al., 2013), the revenue of embedding $vnr_i$ at time $t$ is the sum of all accommodated virtual resources of the virtual network request $vnr_i$ at time $t$.

$$R(vnr_i, t) = Life(vnr_i, t). \left( \sum_{n_{v_i} \in N_{v_i}} CPU(n_{v_i}) + \sum_{l_{v_i} \in L_{v_i}} BW(l_{v_i}) \right) \quad (1)$$

where $CPU(n_{v_i})$ represents the required CPU for the virtual node $n_{v_i}$, $BW(l_{v_i})$ is the required BW for the virtual link $l_{v_i}$, and $Life(vnr_i, t) = 1$ if $vnr_i$ is in its lifetime, otherwise $Life(G_{v_i}, t) = 0$.

*Substrate resources fragmentation (SNF):* substrate resources fragmentation reduces possibilities of accepting new virtual network requests. Although, there are enough available substrate resources to accommodate incoming virtual network requests, virtual network requests are rejected due to spreading of free substrate resources. Substrate resources are fragmented by provisioning and releasing substrate resources over time to accommodate incoming VNRs.

Substrate network $G_s$ is considered fragmented if there are at least two independent sub-substrate networks $G_{s_i}, G_{s_j} \subset G_s$, such that $N_{s_i} \cap N_{s_j} = \emptyset$ and there is no loop-free substrate path $p'_s$ satisfies all connectivity constraints and connects two substrate nodes from $N_{s_i}$ and $N_{s_j}$.

To measure substrate network fragmentation (SNF) at time $t$, we use the following formula:

$$SNF(t) = 1 - \frac{\sum_{i=1}^{m} \left( Residual(G_{s_i}, t) \right)^q}{\left( \sum_{i=1}^{m} Residual(G_{s_i}, t) \right)^q} \quad (2)$$

where $m$ is the number of fragments in the substrate network, q is a positive integer number greater than 1 to reduce the influence of small negligible fragments as long as one large fragment exits, and $Residual(G_{s_i}, t)$ is the total residual substrate resources in sub-substrate network $G_{s_j}$ at time $t$. $Residual(G_{s_i}, t)$ is calculated as following:

$$Residual(G_{s_i}, t) = \sum_{n_{s_i} \in N_{s_i}} CPU_{residual}(n_{s_i}, t) + \sum_{l_{s_i} \in L_{s_i}} BW_{residual}(l_{s_i}, t), \text{ Where } G_{s_i} = (N_{s_i}, L_{s_i})$$

The substrate network fragmentation formula in equation (2) is inspired by the fragmentation measure proposed by Gehr and Schneider in (Gehr & Schneider, 2009).

*Virtual Network Embedding Cost*: as in (Shahin, 2015a; Shahin, 2015b), the cost of embedding $vnr_i$ at time $t$ is defined as total allocated substrate resources to $vnr_i$ at time $t$.

$$Cost(vnr_i, t) = Life(vnr_i, t). \left( \sum_{n_{v_i} \in N_{v_i}} CPU(n_{v_i}) + \sum_{l_{v_i} \in L_{v_i}} BW(l_{v_i}). Length(M_{L_{v_i}}(l_{v_i})) \right) \quad (3)$$

where $Length(M_{L_{v_i}}(l_{v_i}))$ is the length of substrate path that is allocated to the virtual link $l_{v_i}$.

*Objectives*: the main objectives are to increase the revenue of VNE by increasing virtual network acceptance ratio and decreasing the cost of embedding each virtual network. Virtual network acceptance ratio can be increased by decreasing substrate resources fragmentation. The following metrics are used to evaluate and compare VNE algorithms:

- *The long-term average revenue*, which is defined by

$$\lim_{T \to \infty} \left( \frac{\sum_{t=0}^{T} \sum_{i=1}^{I} R(vnr_i, t)}{T} \right) \quad (4)$$

where $I = \| VNR \|$, and $T$ is the total time.

- *The VNR acceptance ratio*, which is defined by

$$\frac{\|VNR_s\|}{\|VNR\|} \quad (5)$$

where $VNR_s$ is the set of all accepted virtual network requests.

- *The long term R/Cost ratio*, which is defined by

$$\lim_{T \to \infty} \left( \frac{\sum_{t=0}^{T} \sum_{i=1}^{I} R(vnr_i, t)}{\sum_{t=0}^{T} \sum_{i=1}^{I} Cost(vnr_i, t)} \right) \quad (6)$$

- *The long-term average substrate network fragmentation*, which is defined by

$$\lim_{T \to \infty} \left( \frac{\sum_{t=0}^{T} SNF(t)}{T} \right) \quad (7)$$

## 5. The proposed MEPDE-VNE algorithm

Algorithm 1 shows the steps of the proposed MEPDE-VNE algorithm. The first step (from line 1 to line 21) is creating the initial population $P(0)$, which is a set of mappings from $G_v$ to $G_s$. To collect these mappings, the MEPDE-VNE algorithm constructs breadth-first searching tree for the $G_v$ graph. The root of the constructed tree is the virtual node with the largest resources. Nodes in each level in the created breadth-first searching tree are sorted in descending order based on their required resources. Candidate substrate nodes list $C_s$ for the root virtual node is created by collecting all substrate nodes that have enough resources to embed the root virtual node. All substrate nodes in $C_s$ are sorted in descending order based on their available resources. MEPDE-VNE algorithm uses $Embed()$ function to collect mappings from $G_v$ to $G_s$. In order to maximize the number of found elements of the Pareto optimal set and to maximize spread of VNE solutions, MEPDE-VNE algorithm visits substrate nodes in the candidate substrate nodes list $C_s$ sequentially and incrementally increases the maximum hop limit in each iteration until the population $P(0)$ is completed. If $Embed()$ function cannot generate the required number of individuals (due to lack of substrate resources), $Population_{Size}$ is reset to the number of generated individuals.

In line 28, each individual in the constructed population is locally improved using $Optimize()$ function. $Optimize()$ function visits virtual nodes and virtual links in each individual in round-robin fashion and uses breadth-first search to find an enhanced remapping. *Optimize()* function terminates when no further improvements can be made.

In line 29, the proposed MEPDE-VNE algorithm uses $Fast\_non\_dominated\_sort()$ function, which applies *Fast Nondominated Sorting* technique proposed in (Deb et al., 2002) to order the constructed population into a hierarchy of non-dominated Pareto fronts and assign a rank value to each individual based on its dominance level and crowding distance value. Crowding-distance values are computed after sorting population according to objective functions in equation 2, and equation 3.

In lines 31-35, the new offspring population $Q(t)$ is generated using $Reproduction()$ function, and optimized using $Optimize()$ function. A combined population $R(t) = Q(t) \cup P(t)$ is created and sorted according to nondomination using $Fast\_non\_dominated\_sort()$ function. To ensure elitism, a new population $P(t + 1)$ is created by retrieving the best nondominated solutions from the combined population $R(t)$ using *Pareto_optimal_front()* function. The new population $P(t + 1)$ of size $Population_{Size}$ is now used for the next iteration. After reaching $Iterations_{Max}$, the best individual in the Pareto optimal front of the final population is returned as a suggested solution.

MEPDE-VNE algorithm uses $Embed()$ function to generate individuals of the initial population. Algorithm 2 shows the steps of the $Embed()$ function. $Embed()$ function creates candidate substrate nodes list, which contains all substrate nodes that have enough CPU to embed current virtual node and have loop free substrate paths with enough bandwidths to embed virtual links between current virtual node and its previously mapped neighbors. The length of each substrate path is less than or equal the maximum substrate path length $Hops$.

ALGORITHM 1: The details of the MEPDE-VNE algorithm

**INPUTS**:
  $G_v = (N_v, L_v)$: VN to be embed
  $G_s = (N_s, L_s)$: SN to embed on
  $Iterations_{Max}$: maximum number of iterations
  $Population_{Size}$: population size
  $Max\_backtrack$: upper bound of nodes re-mapping operation
  $Hops_{Max}$: maximum allowed substrate path length
**OUTPUTS**:
  $M(G_v)$: map VN nodes and links to SN's resources
  $S\_VNE$: VN embedding success flag
**Begin**
  1: Build breadth-first searching tree of $G_v$ from virtual node with largest resources.
  2: Sort all virtual nodes in each level in non-ascending order according to required resources.
  3: Create an empty population $P(t)$ at $t = 0$
  4: $Hops = 0$, where $Hops$ is the maximum allowed substrate path length in current iteration
  5: Build candidate substrate node list $C_s$ for $G_{v_{root}}$
  6: **while** size of $P(t) < Population_{Size}$ **and** $Hops \leq Hops_{Max}$
  7:   **for** each substrate node $n_{s_j} \in C_s$
  8:     Create new map $M'(G_v)$
  9:     $Add\left(\left(n_{v_i}, n_{s_j}\right), M'(G_v)\right)$, where $n_{v_i} = G_{v_{root}}$
  10:    $backtrack\_count = 0$
  11:    **if** $Embed(G_s, n_{v_{i+1}}, P(t), M'(G_v))$ **then**
  12:      $P(t) = P(t) \cup \{M'(G_v)\}$
  13:    **else**
  14:      $Delete\left(\left(n_{v_i}, n_{s_j}\right), M'(G_v)\right)$
  15:    **end if**
  16:    **if** size of $P(t) \geq Population_{Size}$ **then**
  17:      **break**
  18:    **end if**
  19:   **end for**
  20:   $Hops = Hops + 1$
  21: **end while**
  22: **if** size of $P(t) = 0$ **then**
  23:   $S\_VNE = false$
  24:   **return**
  25: **else**
  26:   $Population_{Size} = $ size of $P(t)$
  27: **end if**
  28: $Optimize(P(t))$
  29: $Fast\_non\_dominated\_sort(P(t))$
  30: **while** $t < Iterations_{Max}$
  31:   Create an empty offspring population $Q(t)$
  32:   **while** size of $Q(t) < Population_{Size}$
  33:     $Q(t) = Q(t) \cup Reproduction(P(t))$
  34:   **end while**
  35:   $Optimize(Q(t))$
  36:   $R(t) = Q(t) \cup P(t)$
  37:   $Fast\_non\_dominated\_sort(R(t))$
  38:   $P(t + 1) = Pareto\_optimal\_front(R(t), Population_{Size})$
  39:   $t = t + 1$
  40: **end while**
  41: $M(G_v) = Best\_Pareto\_optimal\_front(P(t))$
  42: $S\_VNE = true$
  43: **return**
**End**

ALGORITHM 2: The details of $Embed()$ Function

**INPUTS:**
  $G_s$: substrate network to embed on
  $n_{v_i}$: current virtual node to be embedded
  $P(t)$: previous population
  $M(G_v)$: map of the previously mapped nodes and links
**OUTPUTS:**
  $M(G_v)$: updated map
  $S_{flag}$: VN embedding success flag
**Begin**
  1: Build candidate substrate node list $C_i$ for $n_{v_i}$
  2: **for** each $n_s$ in $C_i$
  3:   $Add\left(\left(n_{v_i}, n_s\right), M(G_v)\right)$
  4:   **if** $\nexists n_{v_{i+1}}$ **then**
  5:     **if** $M(G_v) \notin P(t)$ **then**
  6:       $S_{flag} = true$
  7:       **return**
  8:     **end if**
  9:   **else**
  10:    **if** $Embed(G_s, n_{v_{i+1}}, P(t), M(G_v))$ **then**
  11:      $S_{flag} = true$
  12:      **return**
  13:    **end if**
  14:  **end if**
  15:  $Delete\left(\left(n_{v_i}, n_s\right), M(G_v)\right)$
  16:  **if** $backtrack\_count > Max\_backtrack$ **then**
  17:    $S_{flag} = false$
  18:    **return**
  19:  **end if**
  20: **end for**
  21: $backtrack\_count ++$
  22: $S_{flag} = false$
  23: **return**
**End**

Current virtual node is sequentially mapped to substrate nodes in its candidate substrate node list. If there is no suitable substrate node in candidate nodes list, we backtrack to the previously mapped node, re-map it to the next candidate node, and continue to the next node. After mapping the last virtual node in the virtual network, if the generated individual already exists in the current population, $Embed()$ function backtrack to the previously mapped node, and re-map it to the next candidate substrate node to generate a new individual. *Reproduction()* function is used to generate new individuals from current population. In this paper, we represent a virtual nodes mapping solution as a chromosome. The number of genes in the chromosome equals to the number of virtual nodes in the virtual network. Each gene corresponds to a virtual node and contains its substrate node.

New individual is generated by performing single-point crossover for two parents, which are picked using *roulette wheel selection* from the current population. Crossover is performed by picking a random midpoint, picking the first part of the nodes mappings from the first parent, and picking the remaining nodes mappings from the second parent. If two connected virtual nodes are picked from the same parent, mapping of the virtual link that connects these two virtual nodes is also picked from the same parent.

If two connected virtual nodes are picked from different parents, virtual link between these virtual nodes is mapped to a shortest loop-free substrate path that satisfies the bandwidth constraint. Although this process does not grantee satisfaction of the CPU and connectivity constraints, *Optimize()* function will remap virtual nodes and virtual links to satisfy the CPU and connectivity constraints and to find the local optimal solution. If *Optimize()* function fails in improving the generated individual, the generated individual will be deleted during creation of the new population $P(t+1)$. The population $P(t+1)$ is created by retrieving best nondominated solutions from the combined population $R(t)$. Remapping process performed by *Optimize()* function maximizes solutions' spreading by moving generated VNE solution in substrate network to find valid solution located between its parents.

After generating new individual, mutation is applied on the new individual. Mutation is performed by remapping virtual node to a randomly selected substrate node from a set of all substrate nodes that are not used in the current population and have enough substrate resources to embed virtual node. Again, virtual links of the mutated virtual node are mapped to shortest loop-free substrate paths that satisfy bandwidth constraints. If no such substrate paths are found, mutation is rolled back. This mutation allows MEPDE-VNE algorithm to investigate new areas in substrate network, so that we can have distributed VNE solutions as smooth and uniform as possible.

## 6. Performance Evaluation

The proposed MEPDE-VNE algorithm is online one stage virtual network embedding algorithm, which embeds virtual nodes with considering the required connectivity constraints. MEPDE-VNE algorithm deals with online version of virtual network embedding problem and does not require any previous knowledge. Virtual links and virtual nodes are embedded at the same stage with high co-ordination to avoid embedding neighboring virtual nodes on widely separated substrate nodes.

Most of current algorithms search one place in solution spaces and return first solution without considering remaining solutions, while MEPDE concurrently searches different parts in the solution space to collect non-dominated solutions and returns best Pareto optimal solution.

One of the main difficulties that faces virtual network embedding algorithms is finding sub-substrate network with topology same as virtual network topology. MEPDE-VNE algorithm overcomes the ossification of the current virtual network embedding algorithms by allowing multiple virtual nodes from the same virtual network to coexist on the same substrate node. Virtual network embedding cost is reduced by eliminating the cost of embedding virtual links between virtual nodes that share same substrate node.

To evaluate the proposed MEPDE-VNE algorithm, we compared its performance with the following algorithms: RW-MaxMatch (Cheng et al., 2011), RW-BFS (Cheng et al., 2011), BFSN-HEM (Shahin, 2015b), vnmFlib (Lischka & Karl, 2009), HCM (Shahin, 2015a), and BFSN (Shahin, 2015b).

*6.1. Evaluation environment settings*

Performance is evaluated using two substrate network topologies, which are generated using Waxman generator. The first SN topology is generated with 50 nodes and 250 links. BW of the substrate links are uniformly distributed between 50 and 100 with average 75. The second SN topology is configured with 200 nodes and 1000 links. BW of the substrate links are uniformly distributed between 50 and 150 with average 100.

Each substrate node is randomly assigned one of the following configurations: HP ProLiant ML110 G4 (Intel Xeon 3040, 2 cores X 1860 MHz, 4 GB), or HP ProLiant ML110 G5 (Intel Xeon 3075, 2 cores X 2660 MHz, 4 GB).

We generated 1000 Virtual network topologies using Waxman generator with connectivity 50%. The number of virtual nodes in each VN is variant from 2 to 20. Each virtual node is randomly assigned one of the following CPU: 2500 MIPS, 2000 MIPS, 1000 MIPS, and 500 MIPS, which are correspond to the CPU of Amazon EC2 instance types. Bandwidths of the virtual links are uniformly distributed between 1 and 50. Arrival times of the virtual network requests are generated randomly with arrival rate 10 VNs per 100 time units. VNR's lifetime are generated randomly between 300 and 700 time units with average 500. Generated SN and VNs topologies are stored in *brite* format and used as inputs for all algorithms. For all algorithms, we set the maximum allowed substrate path length ($Hops_{Max}$) to 2, and the upper bound of backtracking process ($Max\_backtrack$) to $3n$, where $n$ is the number of nodes in each VNR. $Iterations_{Max}$ and $Population_{Size}$ of the MEPDE-VNE algorithm are set to 5 and 10. Finally, we compared the results from the implemented algorithms.

*6.2. Evaluation results*

The long-term average revenue is evaluated using equation 4. Figure 2 shows the long-term average revenue comparison between the proposed algorithm and existing algorithms using the first substrate network (50 nodes and 250 links), and figure 3 shows the long-term average revenue comparison using the second substrate network (200 nodes and 1000 links).

As shown in figure 2 and figure 3, the proposed MEPDE-VNE algorithm increases the revenue among other algorithms. This is expected because most of existing algorithms return the first solution while MEPDE-VNE algorithm tries to get a Pareto optimal set and returns the best one.

Figure 4 and figure 5 show the VNR acceptance ratio, which is defined by equation 5. MEPDE-VNE algorithm increases number of accepted virtual network requests. Although, VNR acceptance ratio defined by equation 5 was used by many researchers to evaluate performance (Cheng et al., 2011; Cheng et al., 2012), it does not consider the VNR resource size. For example, in figure 4, algorithm *vnmFlib* accepted around 29% of the VNRs but it accepted only around 13% of the total VNR resources as shown in figure 6. This variation means that *vnmFlib* algorithm rejected virtual networks with large resources and accepted virtual networks with few resources. MEPDE-VNE algorithm accepted 30% and 80% of the VNRs (figure 4 and figure 5) and accepted 19% and 71% from the total resources of virtual network requests (figure 6 and figure 7). To be more specific, MEPDE-VNE algorithm accepted 19% and 70% of the VNRs' CPU (figure 8 and figure 9) and accepted 12% and 63% of the VNRs' BW (figure 10 and figure 11).

Rejecting virtual network requests is not only due to the lack of substrate resources but also due to the specified maximum number of node remapping operations, which is specified to avoid exponential explosion of node remapping operations. For example, RW-BFS rejected 94% and 92% of the VNR resources (figure 12 and figure 13) while there are 76% and 92% of substrate resources are available (figure 14 and figure 15).

Another reason behind rejecting virtual network requests is preventing more than one virtual node from the same virtual network to be accommodated by the same substrate node. This restriction makes virtual network embedding algorithm try to embed virtual network on sub-substrate network with the same network topology, which is very difficult to find especially in new substrate network topologies (e.g. DCell, DCube, Fat-Tree).

Virtual network requests maybe rejected due to substrate resources fragmentation. MEPDE-VNE algorithm does not fragment substrate resources, as some of existing algorithms do. By using MEPDE-VNE algorithm, long-term average substrate network fragmentation, which is defined by equation 7, stays zero while some of existing algorithms fragment substrate resources as shown in figure 16.

Figure 17 and figure 18 show comparison of long-term R/Cost ratio, which is defined by equation 6. In both figures, long-term R/Cost ratio of the MEPDE-VNE algorithm exceeds 100% because cost of embedding virtual networks is less than the revenue gained from embedding them. Figures 19-22 compare virtual network embedding time consumed by different VNE algorithms. MEPDE-VNE algorithm consumes more time to find the best Pareto optimal solution due to the large number of solutions. As shown in figure 23 and figure 24, MEPDE-VNE algorithm increases substrate resources utilization by accommodating virtual resources as much as possible. Figures 25-28 show substrate resources utilization in more details.

Figure 12 and figure 13 show that MEPDE-VNE algorithm rejected 80% and 29% of incoming virtual resources (figures 29-32 show details of rejected virtual resources), but this is due to the lack of available substrate CPU (figures 33-36 show details of available substrate CPU). Finally, figure 37 and figure 38 compare number of active substrate nodes that are used to achieve virtual network requests.

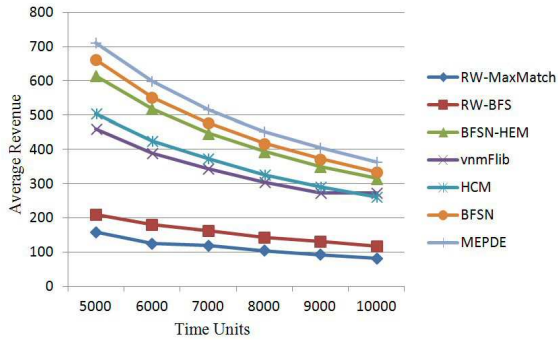

Figure 2. Revenue comparison using 50 substrate

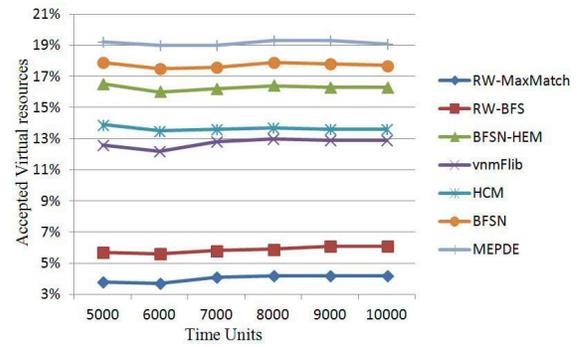

Figure 6. Virtual resources acceptance ratio comparison using 50 substrate nodes

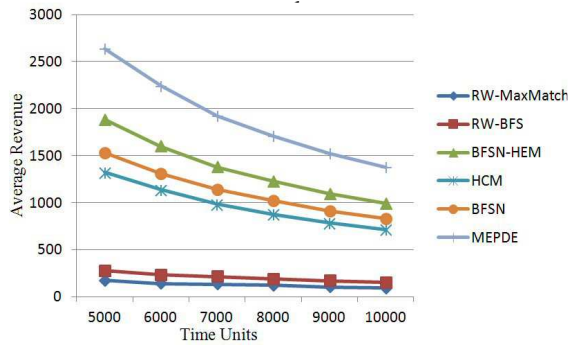

Figure 3. Revenue comparison using 200 substrate node

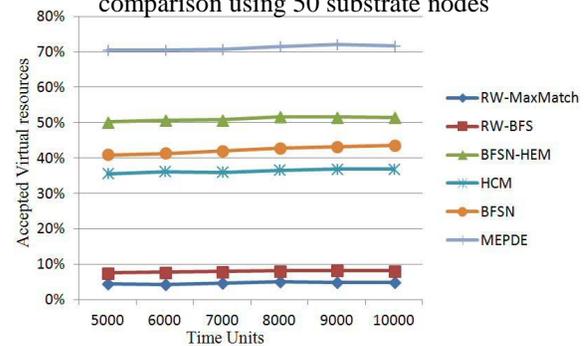

Figure 7. Virtual resources acceptance ratio comparison using 200 substrate nodes

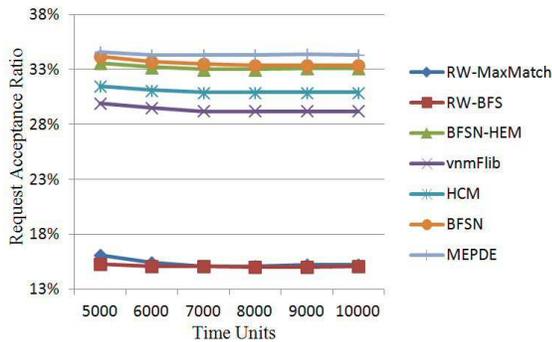

Figure 4. VNR acceptance ratio comparison using 50 substrate nodes

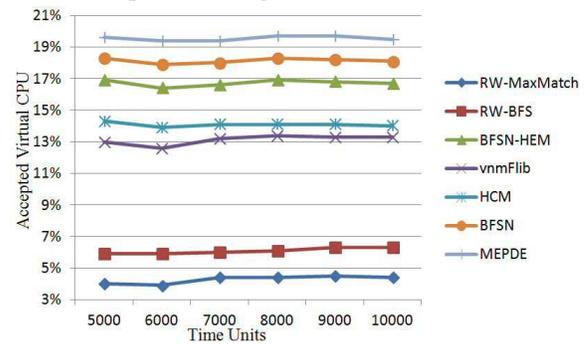

Figure 8. Accepted virtual CPU comparison using 50 substrate nodes

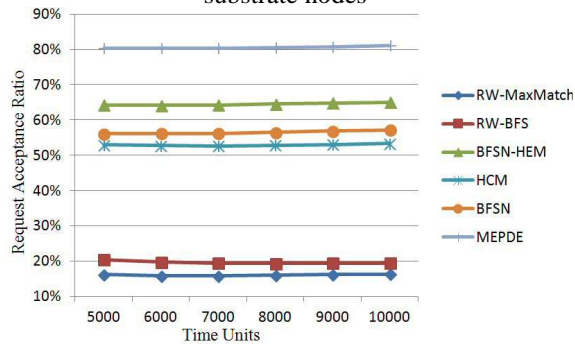

Figure 5. VNR acceptance ratio comparison using 200 substrate nodes

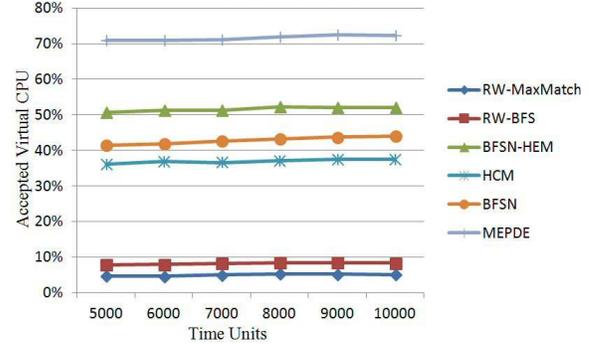

Figure 9. Accepted virtual CPU comparison using 200 substrate nodes

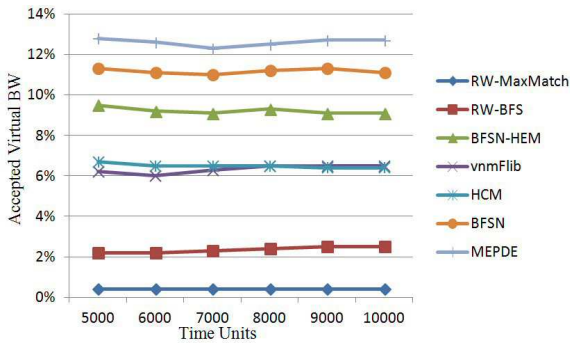

Figure 10. Accepted virtual BW comparison using 50 substrate nodes

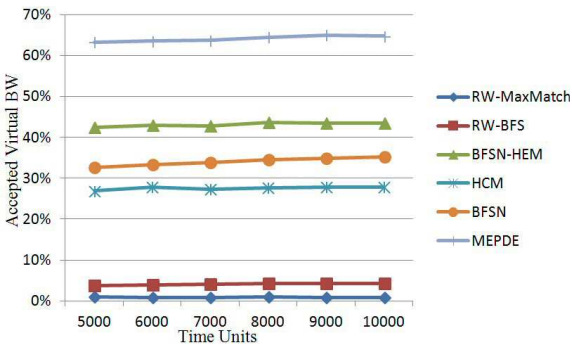

Figure 11. Accepted virtual BW comparison using 200 substrate nodes

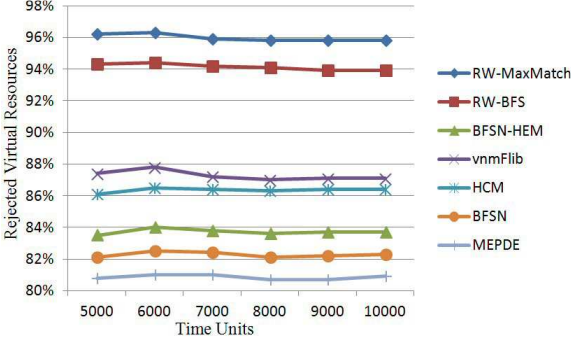

Figure 12. Rejected virtual resources comparison using 50 substrate nodes

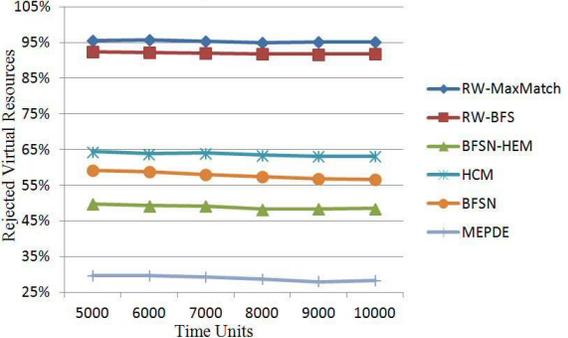

Figure 13. Rejected virtual resources comparison using 200 substrate nodes

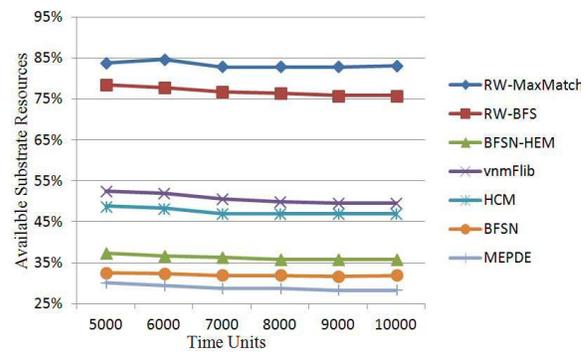

Figure 14. Available substrate resources comparison using 50 substrate nodes

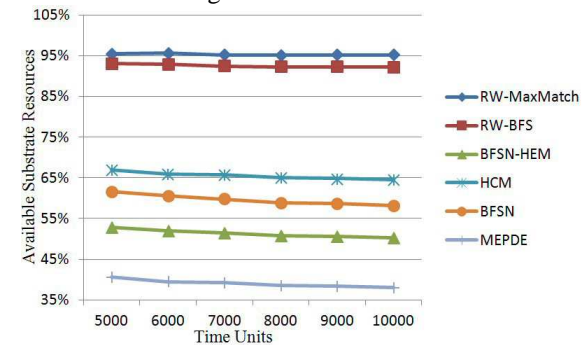

Figure 15. Available substrate resources comparison using 200 substrate nodes

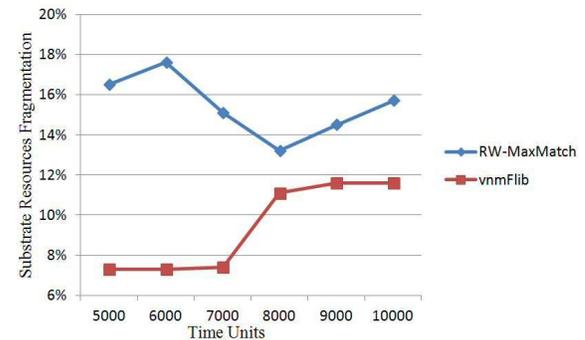

Figure 16. Substrate resources fragmentation comparison using 50 substrate nodes

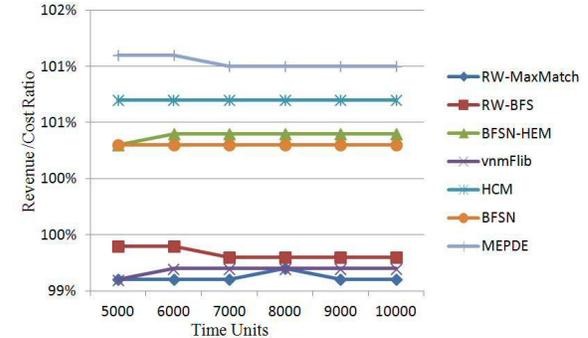

Figure 17. Revenue/Cost comparison using 50 substrate nodes

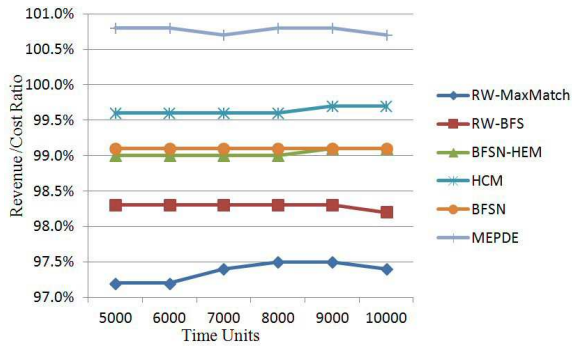

Figure 18. Revenue/Cost comparison using 200 substrate nodes

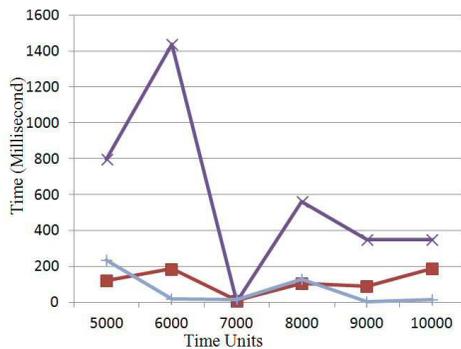

Figure 19. Virtual network embedding time comparison between RW-BFS, vnmFlib, and MEPDE algorithms using 50 substrate nodes

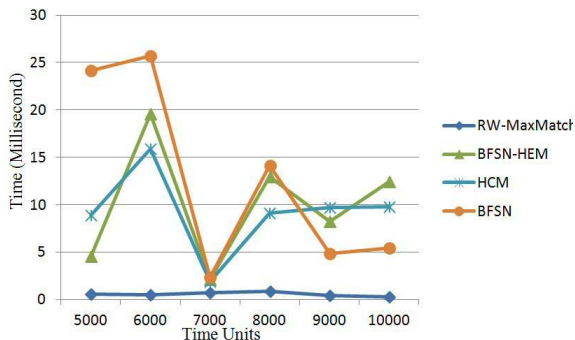

Figure 20. Virtual network embedding time comparison between RW-MaxMatch, BFSN-HEM, HCM, and BFSN algorithms using 50 substrate nodes

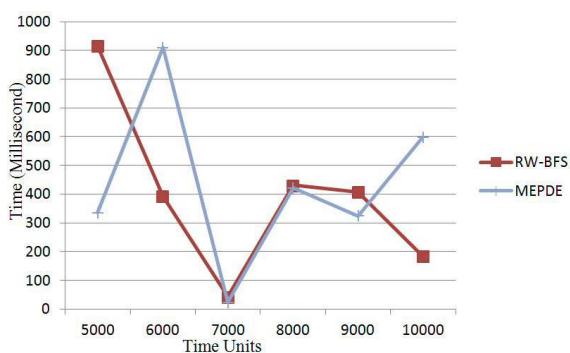

Figure 21. Virtual network embedding time comparison between RW-BFS, and MEPDE algorithms using 200 substrate nodes

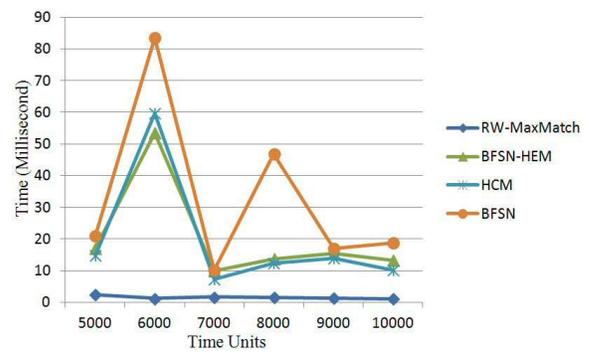

Figure 22. Virtual network embedding time comparison between RW-MaxMatch, BFSN-HEM, HCM, and BFSN algorithms using 200 substrate

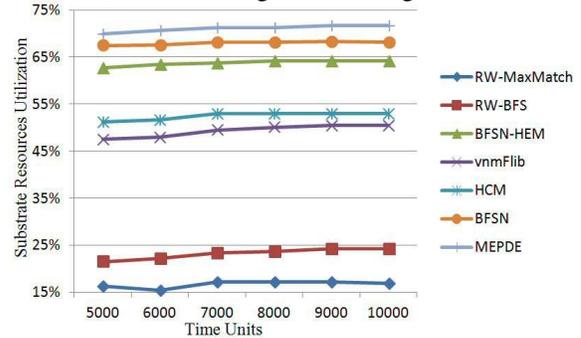

Figure 23. Substrate resources utilization comparison using 50 substrate nodes

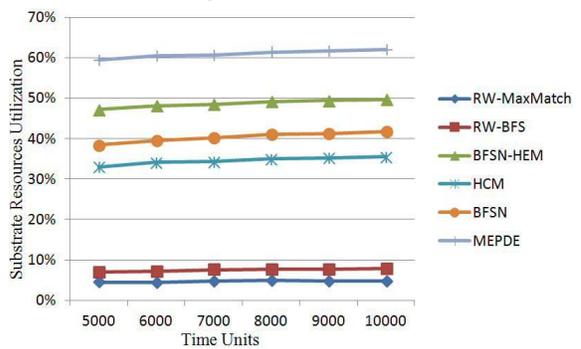

Figure 24. Substrate resources utilization comparison using 200 substrate nodes

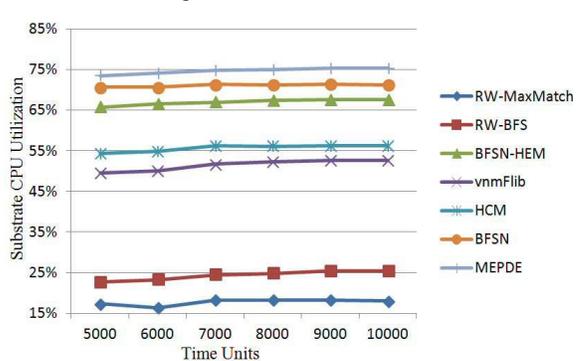

Figure 25. Substrate CPU utilization comparison using 50 substrate nodes

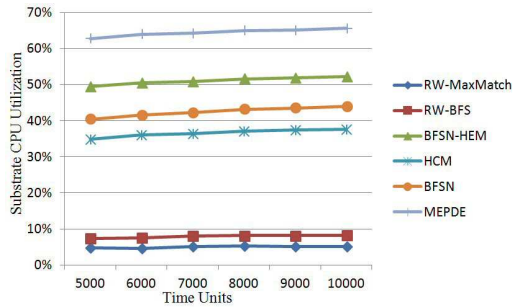

Figure 26. Substrate CPU utilization comparison using 200 substrate nodes

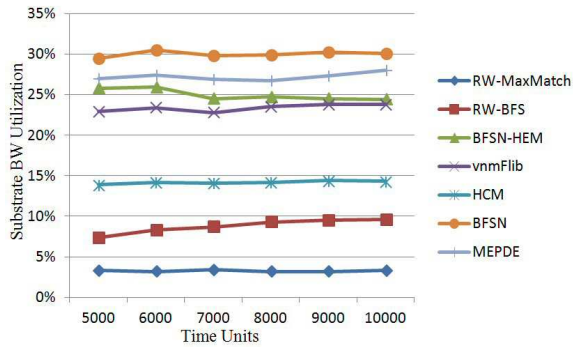

Figure 27. Substrate BW utilization comparison using 50 substrate nodes

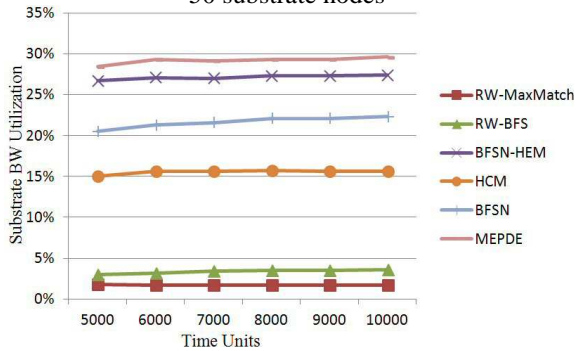

Figure 28. Substrate BW utilization comparison using 200 substrate nodes

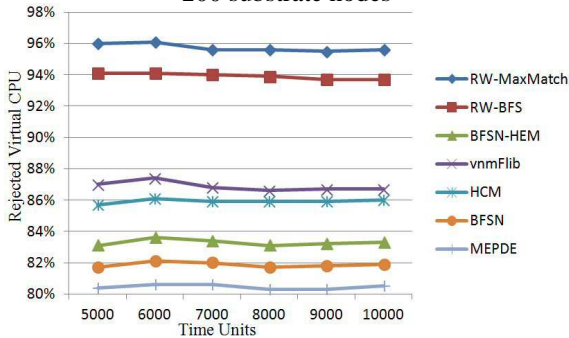

Figure 29. Rejected virtual CPU comparison using 50 substrate nodes

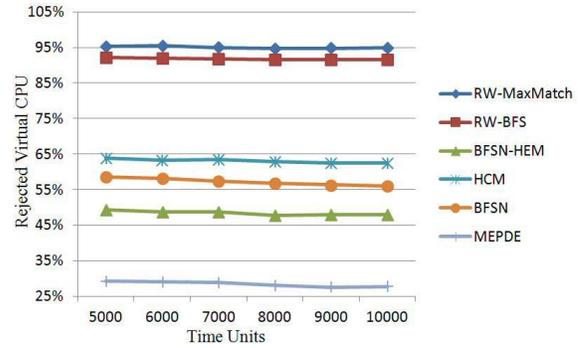

Figure 30. Rejected virtual CPU comparison using 200 substrate nodes

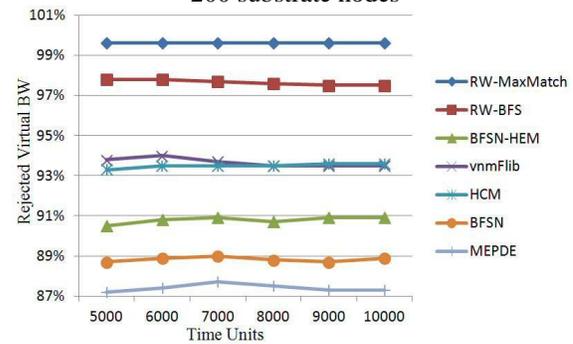

Figure 31. Rejected virtual BW comparison using 50 substrate nodes

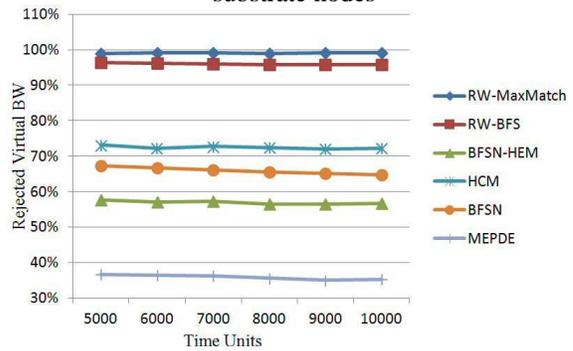

Figure 32. Rejected virtual BW comparison using 200 substrate nodes

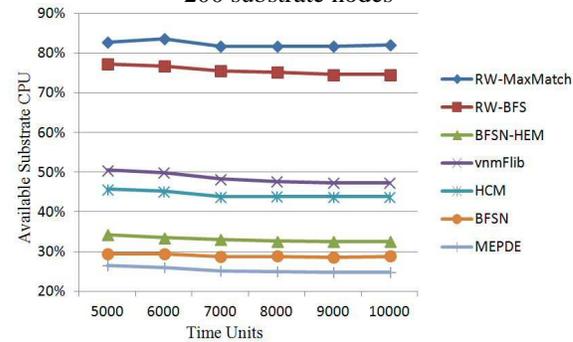

Figure 33. Available substrate CPU comparison using 50 substrate nodes

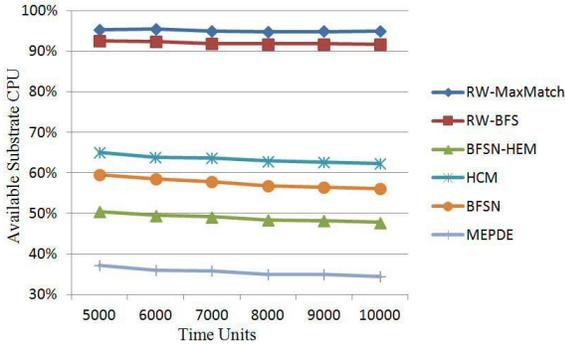

Figure 34. Available substrate CPU comparison using 200 substrate nodes

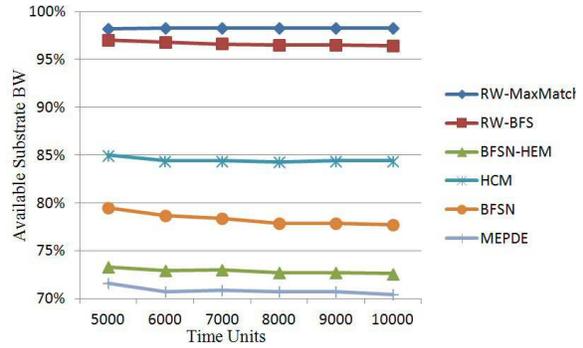

Figure 36. Available substrate BW comparison using 200 substrate nodes

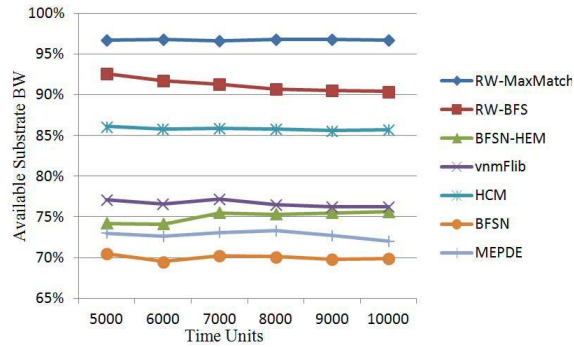

Figure 35. Available substrate BW comparison using 50 substrate nodes

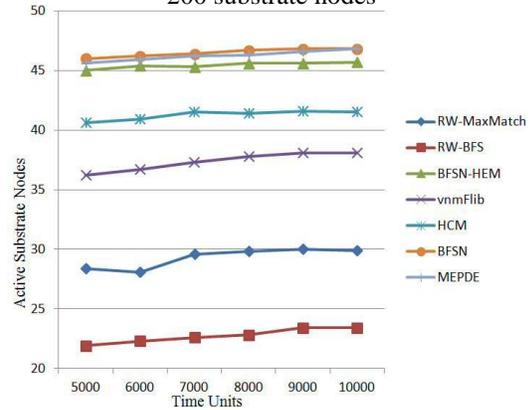

Figure 37. Active substrate nodes comparison using 50 substrate nodes

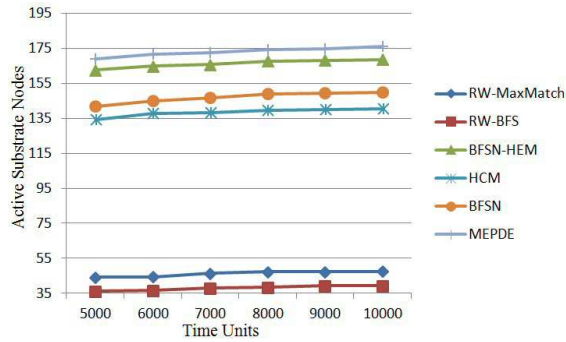

Figure 38. Active substrate nodes comparison using 200 substrate nodes

## 7. Conclusion

Virtual network embedding problem is one of the major challenges that cloud-computing platforms currently face. In the last few years, several optimal virtual network-embedding algorithms have been proposed to find optimal solutions in small instances. However, finding optimal virtual network embedding solution using traditional virtual network embedding algorithms in large-scale cloud computing platforms becomes unaffordable. This observation motivates us to propose memetic elitist pareto evolutionary algorithm for virtual network embedding, called MEPE-VNE. MEPE-VNE algorithm increases the acceptance ratio of virtual network requests by minimizing substrate resources fragmentation. The proposed algorithm reduces virtual network embedding cost by increasing number of virtual links that are achieved by inter ram switch techniques (virtual links between virtual nodes that coexist in the same substrate node). MEPE-VNE provides near optimal virtual network embedding solution in reasonable time in large-scale data centers by simultaneously explores different parts in Pareto optimal set. Each solution in the Pareto optimal solutions is enhanced using local search algorithm to increase Pareto optimal set quality. Non-dominated

Pareto fronts are sorted using non-dominated sorting genetic algorithm-II (NSGA-II) after assigning a rank value to each individual based on its dominance level and crowding distance value. MEPE-VNE passes solutions with best fitness values to the offspring population to prevent random destruction of best solutions by crossover or mutation operators. Extensive simulations show that our algorithm outperforms existing algorithms in terms of the long-term average revenue and the acceptance ratio of the virtual network requests. In our future work, we intend to extend the proposed algorithm to solve virtual network embedding problem in distributed cloud computing platforms. In addition, migration of virtual nodes and virtual links will be employed to minimize the rejecting rate of virtual network requests.